\documentclass[11pt, a4paper]{article}
\usepackage{fullpage}
\usepackage{amsfonts}
\usepackage{amssymb}
\usepackage{amsmath}
\usepackage{amsthm}
\usepackage{graphicx}
\usepackage{bm}
\usepackage[makeroom]{cancel}
\usepackage{enumitem}
\usepackage{url}
\usepackage[margin=.9in]{geometry}
\usepackage{amsopn}
\usepackage{mathtools}
\usepackage{hyperref}
\usepackage{doi}
\usepackage{cite}
\usepackage[symbol]{footmisc}

\DeclareMathOperator*{\argmin}{arg\,min}
\newcommand{\R}{\mathbb{R}}
\newcommand{\C}{\mathbb{C}}
\newcommand*{\vertbar}{\rule[-1ex]{0.5pt}{2.5ex}}

\usepackage{listings}
\usepackage{xcolor}
\definecolor{codegreen}{rgb}{0,0.6,0}
\definecolor{codegray}{rgb}{0.5,0.5,0.5}
\definecolor{codepurple}{rgb}{0.58,0,0.82}
\definecolor{backcolour}{rgb}{0.95,0.95,0.92}
\lstdefinestyle{mystyle}{
    backgroundcolor=\color{backcolour},   
    commentstyle=\color{codegreen},
    keywordstyle=\color{magenta},
    numberstyle=\tiny\color{codegray},
    stringstyle=\color{codepurple},
    basicstyle=\ttfamily\normalsize,
    breakatwhitespace=false,         
    breaklines=true,                 
    captionpos=b,                    
    keepspaces=true,                 
    numbers=left,                    
    numbersep=3pt,                  
    showspaces=false,                
    showstringspaces=false,
    showtabs=false,                  
    tabsize=4,
}
\begin{document}
\begin{center}
    \LARGE \bf PyDMD: A Python package for robust dynamic mode decomposition
\end{center}
\begin{center}
    Sara M. Ichinaga$^1$\footnote[1]{Corresponding authors (sarami7@uw.edu)},
    Francesco Andreuzzi$^{2, 6}$,
    Nicola Demo$^{2}$,
    Marco Tezzele$^{3}$,
    Karl Lapo$^{4}$, \\
    Gianluigi Rozza$^{2}$,
    Steven L. Brunton$^{5}$,
    J. Nathan Kutz$^{1}$
\end{center}
\begin{center}
    \scriptsize{
    ${}^1$ Department of Applied Mathematics, University of Washington, Seattle, WA 98195, United States \\ 
    ${}^2$ Mathematics Area, mathLab, SISSA, via Bonomea 265, I-34136 Trieste, Italy \\ 
    ${}^3$ Oden Institute for Computational Engineering and Sciences, University of Texas at Austin, Austin, TX 78712, United States \\
    ${}^4$ Department of Atmospheric and Cryospheric Sciences, University of Innsbruck, Innrain 52, 6020 Innsbruck, Austria \\
    ${}^5$ Department of Mechanical Engineering, University of Washington, Seattle, WA 98195, United States \\ 
    ${}^6$ CERN, Geneva, Switzerland
    }
\end{center}

\begin{abstract}
The \textit{dynamic mode decomposition} (DMD) is a simple and powerful data-driven modeling technique that is capable of revealing coherent spatiotemporal patterns from data. The method's linear algebra-based formulation additionally allows for a variety of optimizations and extensions that make the algorithm practical and viable for real-world data analysis. As a result, DMD has grown to become a leading method for dynamical system analysis across multiple scientific disciplines. \texttt{PyDMD} is a Python package that implements DMD and several of its major variants. In this work, we expand the \texttt{PyDMD} package to include a number of cutting-edge DMD methods and tools specifically designed to handle dynamics that are noisy, multiscale, parameterized, prohibitively high-dimensional, or even strongly nonlinear. We provide a complete overview of the features available in \texttt{PyDMD} as of version 1.0, along with a brief overview of the theory behind the DMD algorithm, information for developers, tips regarding practical DMD usage, and introductory coding examples. All code is available at \href{https://github.com/PyDMD/PyDMD}{\texttt{https://github.com/PyDMD/PyDMD}}.
\end{abstract}

\section{Introduction}
In recent years, the availability and abundance of high-fidelity data across the sciences has greatly motivated the utilization of, as well as the necessity for, algorithms that are accurate, efficient, intuitive, and purely data-driven. One algorithm that has recently emerged as a powerful method for analyzing dynamical system data is the \textit{dynamic mode decomposition} (DMD)~\cite{schmid_2010,tu_2014,dmd_book,schmid2022dynamic}.  DMD generally seeks a low-dimensional set of key spatiotemporal modes that describe a set of observations. This information then allows for a variety of tasks, including dimensionality reduction, state reconstruction, future-state prediction, and system control \cite{dmd_book, data_book}. Hence, despite its conception as a method for analyzing fluid flows \cite{schmid_2009, schmid_2010, schmid_2008, Noack2016jfm}, DMD has since been applied to data sets spanning multiple scientific disciplines \cite{proctor_2015, brunton_2016, alfatlawi_2020, mann_2016, taylor_2018, kaptanoglu_2020, grosek_2014, berger_2015, abraham_2019, bruder_2019, sinha_2020, susuki_1, susuki_2}, and has become the standard approach for approximating the Koopman operator from data \cite{rowley_2009, tu_2014, modern_koopman}. It thus remains imperative that DMD and its growing suite of innovations and algorithms remain both intuitive and accessible for scientists and engineers with diverse backgrounds in mathematics.

\begin{figure}
    \centering
    \includegraphics[scale=0.64]{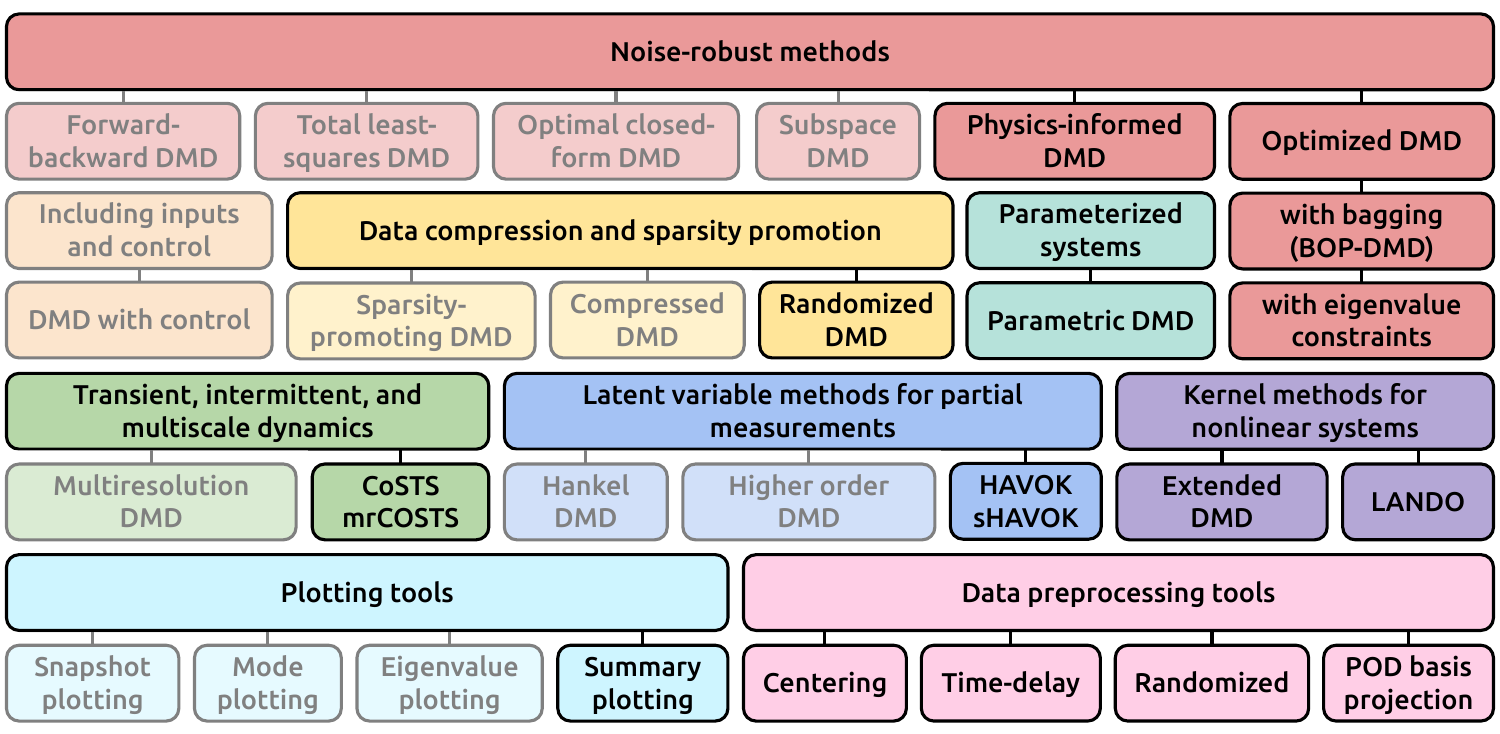}
    \caption{Summary of all \texttt{PyDMD} functionalities as of version 1.0. All features are organized according to use case, and features introduced to \texttt{PyDMD} as a part of this update are indicated via vivid boxes. Previously-available features are represented via semi-transparent boxes.}
    \label{fig:pydmd-overview}
\end{figure}

\texttt{PyDMD} is a Python package that provides the tools necessary for executing the DMD pipeline within an abstract user-friendly interface. Initially released in 2018, the original \texttt{PyDMD} package \cite{pydmd} implemented a wide variety of DMD algorithms \cite{dmdc, mrdmd, spdmd, cdmd, hankeldmd, hodmd, fbdmd, tdmd, not_optdmd, subspacedmd}, all of which we summarize in Figure~\ref{fig:pydmd-overview}. However since then, many crucial DMD variants have arisen, such as optimized DMD for optimal noise suppression \cite{optdmd, bopdmd}, coherent spatiotemporal scale separation (CoSTS) for multiscale measurements \cite{sliding_dmd}, parametric DMD for parameterized systems \cite{paradmd, paradmd_2}, randomized DMD for data compression \cite{rdmd}, and physics-informed DMD for enforcing DMD model constraints \cite{pidmd}. The package also initially lacked tools for analyzing highly nonlinear systems \cite{edmd, edmd_kern, edmd_2, havok, shavok, lando}, as well as general data preprocessors that are often necessary for successful DMD analyses \cite{centering, takens, embedology}. All of this has motivated our recent work to incorporate these powerful DMD variants and features into \texttt{PyDMD}, which we also summarize in Figure \ref{fig:pydmd-overview}.

With this new update to \texttt{PyDMD}, users gain access to state-of-the-art DMD methods, algorithms, and tools that are specifically geared towards real-world data modeling scenarios. Our code is tested, open-source, thoroughly-documented, supplemented with a large suite of \href{https://github.com/PyDMD/PyDMD/tree/master/tutorials}{{\color{blue}Jupyter Notebook tutorials}}, and modularly-structured to allow for future contributions and extensions of the package.

\section{Mathematical Background} \label{sec:background}
Given snapshots $\mathbf{x}(t) \in \R^n$ collected at times $\{t_k\}_{k=1}^m$ and organized into the columns of the matrix
\begin{equation}
\label{eq:X}
\mathbf{X} = 
\begin{bmatrix}
\vertbar & \vertbar & & \vertbar \\
\mathbf{x}(t_1) &
\mathbf{x}(t_2) & 
\dots & 
\mathbf{x}(t_m) \\
\vertbar & \vertbar & & \vertbar
\end{bmatrix} \in \R^{n \times m},
\end{equation}
the DMD algorithm seeks a rank-$r$ spatiotemporal decomposition of $\mathbf{X}$ with the following form:
\begin{equation}
\label{eq:dmd}
\mathbf{X}
\approx 
\begin{bmatrix}
    \vertbar & & \vertbar \\
    \bm{\phi}_1 & \dots & \bm{\phi}_r \\ 
    \vertbar & & \vertbar
\end{bmatrix}
\begin{bmatrix}
    b_1 \\ & \ddots \\ && b_r
\end{bmatrix}
\begin{bmatrix}
    e^{\omega_1 t_1} & \dots & e^{\omega_1 t_m} \\ 
    \vdots & \ddots & \vdots \\
    e^{\omega_r t_1} & \dots & e^{\omega_r t_m}
\end{bmatrix}
= \bm{\Phi} \text{diag}(\mathbf{b})\mathbf{T}(\bm{\omega}).
\end{equation}
With $\bm{\phi}_j \in \C^n$ we denote the $j$th dominant spatial mode of the system, where $\omega_j \in \C$ captures the temporal behavior of $\bm{\phi}_j$. Each $b_j \in \C$ thus represents an appropriate amplitude for the $j$th spatiotemporal mode for accurate system reconstructions.

Although each DMD algorithm seeks the diagnostics given by $\mathbf{\Phi}, \bm{\omega}$, $\mathbf{b}$, variants of the algorithm differ in the way that these are computed. For example, the \textit{exact DMD} algorithm \cite{tu_2014} seeks the eigendecomposition of the linear operator $\mathbf{A} \in \R^{n \times n}$ that best advances the snapshot data one step forward in time, as one may obtain $\mathbf{\Phi}$ and $\bm{\omega}$ from the eigenvectors and eigenvalues of $\mathbf{A}$ respectively. This is done by defining a second data matrix $\mathbf{X}' \in \R^{n \times m}$ similar to \eqref{eq:X}, with each column advanced one time step $\Delta t$ into the future.  It then follows that $\mathbf{A}$ must satisfy
\begin{equation}
\label{eq:linear}
\mathbf{X}' \approx \mathbf{A} \mathbf{X}.
\end{equation}
In practice, the eigendecomposition of $\mathbf{A}$ is recovered from that of a rank-$r$ approximation $\Tilde{\mathbf{A}} \in \R^{r \times r}$ of $\mathbf{A}$ for $r \ll n$, as it can be prohibitively expensive to explicitly compute and decompose $\mathbf{A}$ for large $n$. Once obtained, the eigenvectors $\phi_j$ of the matrix $\mathbf{A}$ give the spatial modes in \eqref{eq:dmd}, while the eigenvalues $\lambda_j$ of $\mathbf{A}$ give the entries of $\bm{\omega}$ via the relationship $\omega_j = \log(\lambda_j) / \Delta t$. One may then obtain the amplitudes $\mathbf{b}$ via Equation \eqref{eq:dmd}. The simplest and most common approach involves fitting to the initial condition with $\mathbf{b} = \mathbf{\Phi}^\dagger \mathbf{x}(t_1)$, though more sophisticated strategies for computing $\mathbf{b}$ have been developed \cite{computing_b, spdmd, optdmd}. For more details on the exact DMD algorithm, we refer readers to \cite{dmd_book}.

Since its conception, the exact DMD algorithm has been widely adapted and modified to improve its breadth and robustness. However, one of the most significant drawbacks of exact DMD is its sensitivity to measurement noise \cite{bagheri_2014, noise_1, noise_2, fbdmd, tdmd}, and in response, a number of noise-robust variants have been developed \cite{fbdmd, consistent_dmd, tdmd, not_optdmd, subspacedmd, optdmd, bopdmd, pidmd}. One of the most recent and effective of these noise-robust approaches is optimized DMD \cite{optdmd}, which uses variable projection for nonlinear least squares problems in order to solve \eqref{eq:dmd} directly via the following nonlinear optimization problem:
\begin{equation}
\label{eq:optdmd}
    \mathbf{\Phi} \text{diag}(\mathbf{b}), \: \bm{\omega}
    = \argmin_{\mathbf{\Phi}_b, \: \bm{\omega}}
    \| \mathbf{X} - \mathbf{\Phi}_b \mathbf{T}(\bm{\omega}) \|_F.
\end{equation}
This approach to DMD has several advantages, with the most prominent being the method's ability to optimally suppress the effects of noise and its ability to handle snapshots that are unevenly sampled in time. The use of variable projection additionally permits the application of constraints and regularizers to the computed DMD diagnostics for added customization and robustness to noise. Sashidhar and Kutz \cite{bopdmd} showed that the results of \eqref{eq:optdmd} can be stabilized and improved through the use of statistical bagging techniques. This result gave rise to \textit{bagging, optimized DMD} (BOP-DMD), which we note is a state-of-the-art and generally recommended all-purpose DMD method for noisy data, as of the writing of this paper.  BOP-DMD is also related to the spectral POD from the field of fluid mechanics~\cite{Towne2018jfm}. 

\begin{figure}
    \centering
    \includegraphics[scale=0.645]{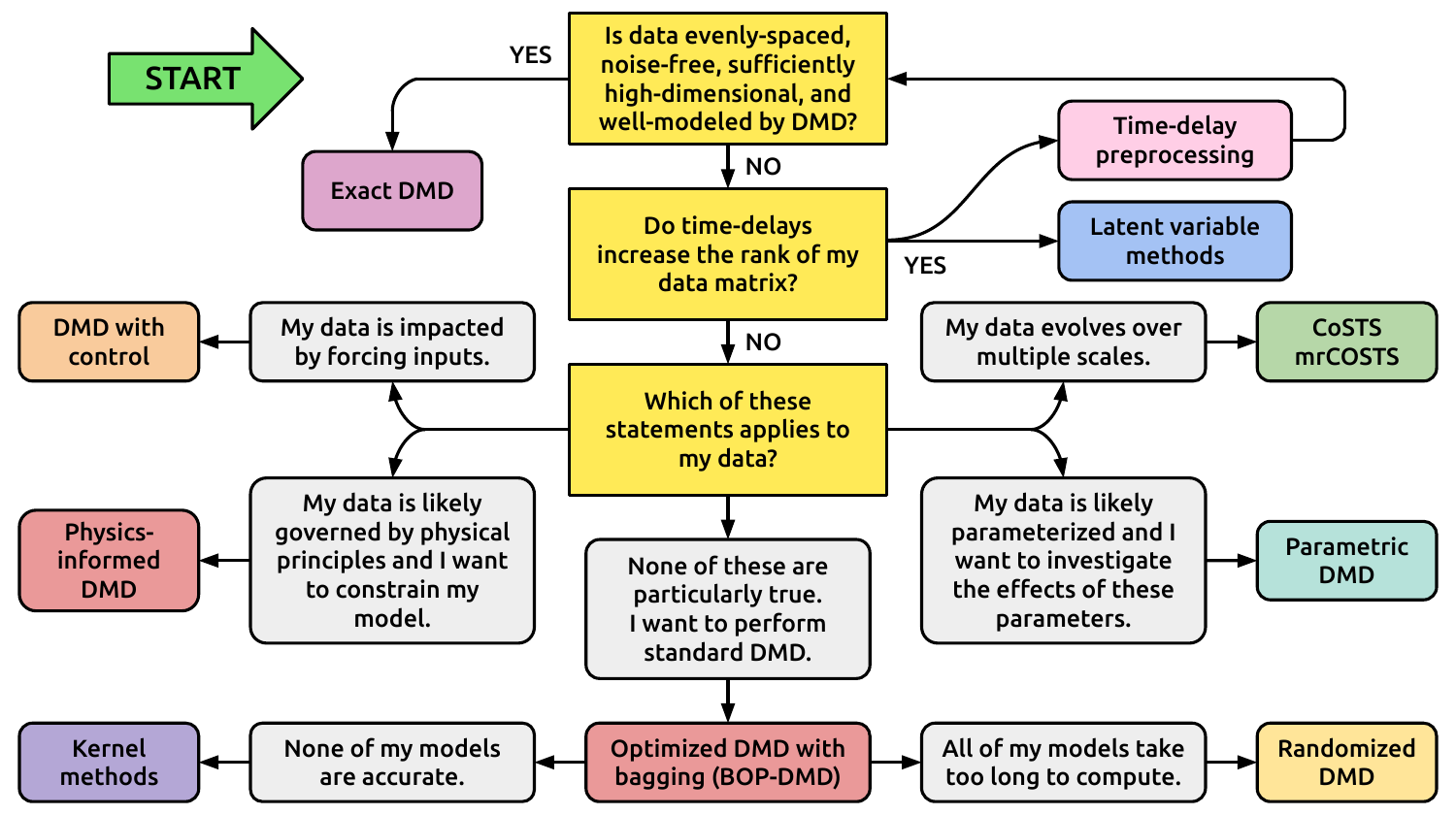}
    \caption{Flow chart for determining an appropriate DMD method based on problem type and data set. DMD methods and tools are color-coded following Figure \ref{fig:pydmd-overview}.}
    \label{fig:dmd-flow}
\end{figure}

Several DMD variants alternatively reformulate the standard DMD algorithm in order to handle systems that cannot be modeled by Equation \eqref{eq:linear}. Indeed, methodological extensions have arisen in order to address systems with inputs and control \cite{dmdc, folkestad2020extended}, systems that exhibit multiscale dynamics \cite{mrdmd, sliding_dmd}, systems that are parameterized \cite{paradmd, paradmd_2}, and even highly nonlinear systems that cannot be modeled globally with a linear operator \cite{edmd, edmd_kern, edmd_2, lando}. Several methods \cite{havok, hankeldmd, hodmd, shavok} use time-delay coordinates in order to reveal and utilize hidden or latent variables from the available data. This approach is rooted in well-established time-delay embedding theory \cite{takens, embedology} which allows us to apply DMD even when we only have access to partial measurements, i.e. data that lacks all relevant system states. Other notable DMD variants include randomized DMD \cite{rdmd}, which uses randomized linear algebra techniques to drastically improve runtime, and physics-informed DMD \cite{pidmd}, which similar to the measure-preserving extended DMD approach \cite{colbrook2023mpedmd} constrains the structure of $\mathbf{A}$ in Equation \eqref{eq:linear} to enforce physical principles and improve robustness to noise. Tensorized formulations of the DMD algorithm have also been developed \cite{klus2018tensor}, along with several more recent robust methods \cite{colbrook2023residual, colbrook2023mpedmd, colbrook2024rigorous} which serve as strong candidates for future extensions of the \texttt{PyDMD} package.

With that being said, it is crucial to consider the nature of one's data and to monitor the accuracy of one's models when applying DMD in practice, as certain variants lend themselves to specific problems, data sets, and shortcomings of exact DMD. We summarize how one might choose an appropriate DMD variant in Figure \ref{fig:dmd-flow}.

\section{PyDMD Structure}
\label{sec:code}

\begin{figure}
    \centering
    \includegraphics[scale=0.64]{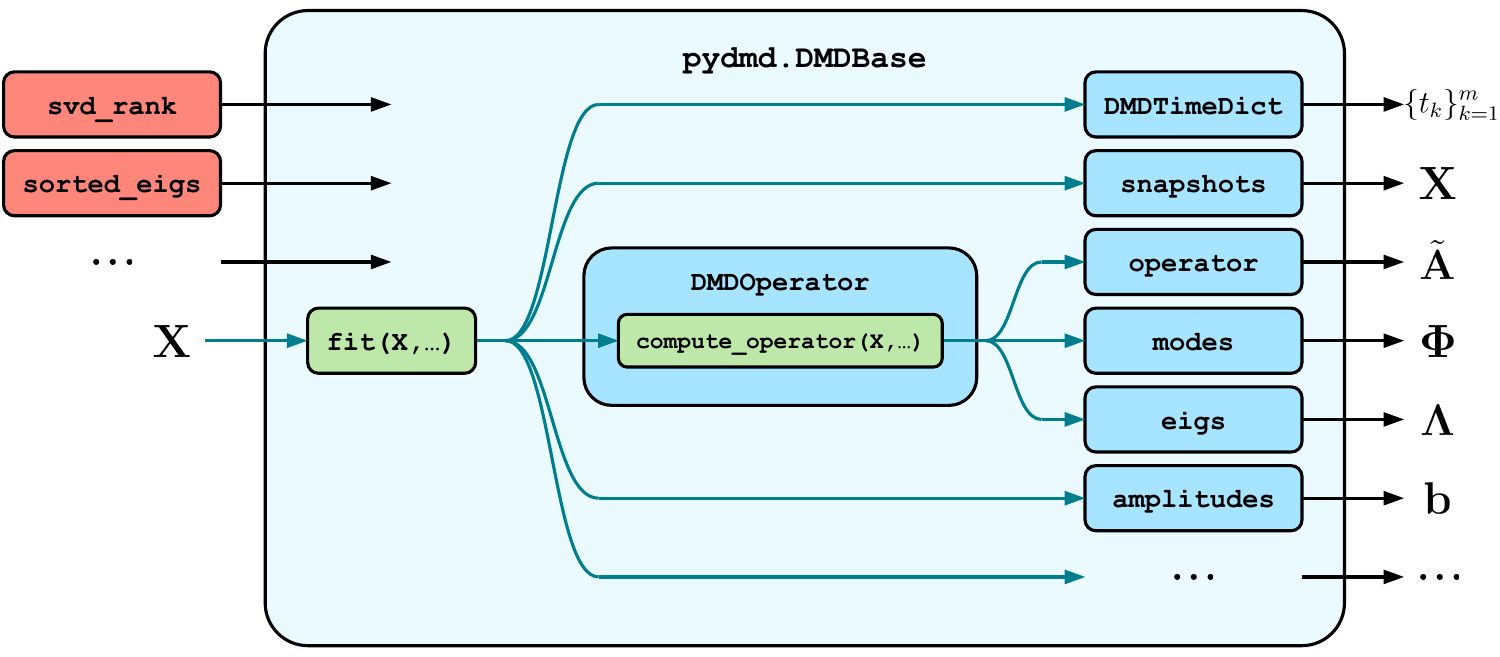}
    \caption{Schematic of a typical \texttt{PyDMD} module. In general, modules keep track of user-inputted parameters, a DMD operator, DMD diagnostics, and information on the input data, among other attributes. Modules also implement a \texttt{fit} method, which when called and given data, performs the DMD pipeline and makes the results available for user access. Note that each module of \texttt{PyDMD} implements a unique DMD variant, where any number of these steps may be performed differently.}
    \label{fig:module}
\end{figure}

The \texttt{PyDMD} package is modular, with most DMD variants possessing their own module within the package. The \texttt{DMDBase} class forms a foundation for the vast majority of modules, as it implements a variety of functionalities that are universal across most DMD variants. Hence in order to avoid code duplication, most modules are implemented by inheriting the \texttt{DMDBase}. It must be said however that several DMD variants differ greatly from the exact DMD algorithm. In this case, it is preferred to implement such techniques from scratch in order to ensure that the final class contains the needed members. In general, all \texttt{PyDMD} modules are capable of the following:
\begin{itemize}
    \item \texttt{PyDMD} modules accept and store parameters of the DMD pipeline. This may include the rank $r$ of the decomposition or any number of attributes. \texttt{DMDBase} defines several commonly-used parameters, though most modules define and use parameters beyond those of the base class.

    \item \texttt{PyDMD} modules sometimes keep track of a \texttt{DMDOperator} and \texttt{DMDTimeDict}s. The \texttt{DMDOperator} represents the reduced operator $\Tilde{\mathbf{A}}$ and keeps track of the eigenvectors $\mathbf{\Phi}$ and eigenvalues $\mathbf{\Lambda}$ of the full operator $\mathbf{A}$. It also computes $\Tilde{\mathbf{A}}$ given data via its \texttt{compute\_operator} method. The \texttt{DMDTimeDict}s store the times of the input snapshots and the times of system reconstruction.

    \item \texttt{PyDMD} modules must implement a \texttt{fit} method, which takes the data matrix $\mathbf{X}$, and possibly some additional input data, and performs DMD. This method typically (1) prepares and stores the input data, (2) calls the \texttt{DMDOperator}'s \texttt{compute\_operator} function, (3) sets the \texttt{DMDTimeDict}s, and (4) uses the operator properties to compute the amplitudes. For less conventional modules, it suffices that \texttt{fit} simply computes and stores crucial DMD information.

    \item \texttt{PyDMD} modules, once fitted, are able to store and fetch the computed DMD diagnostics. They can also use these diagnostics for various tasks, such as system reconstruction and prediction. Extensions of the base class are free to rewrite these functionalities, as well as add new ones.
\end{itemize}
We visualize this general module structure in Figure \ref{fig:module}. We also direct potential developers to the official \href{https://pydmd.github.io/PyDMD/index.html}{{\color{blue}\texttt{PyDMD} documentation}} for an exhaustive description of the parameters and functionalities of the \texttt{DMDBase} class, as well as to our \href{https://github.com/PyDMD/PyDMD/tree/master/tutorials}{{\color{blue}developer tutorials}} for more information on how to develop new modules and future contributions to the \texttt{PyDMD} package.

\section{Examples}
\label{sec:examples}
In this section, we analyze a simple synthetic data set with \texttt{PyDMD} in order to showcase some of the most basic and crucial tools of the package. For more examples and for more tutorials that highlight specific DMD variants and use-cases, see our complete set of \href{https://github.com/PyDMD/PyDMD/tree/master/tutorials}{{\color{blue}Jupyter Notebook tutorials}}.

\subsection{Basic PyDMD usage}
\label{sec:example-1}
\begin{figure}
    \centering
    \includegraphics[scale=0.465]{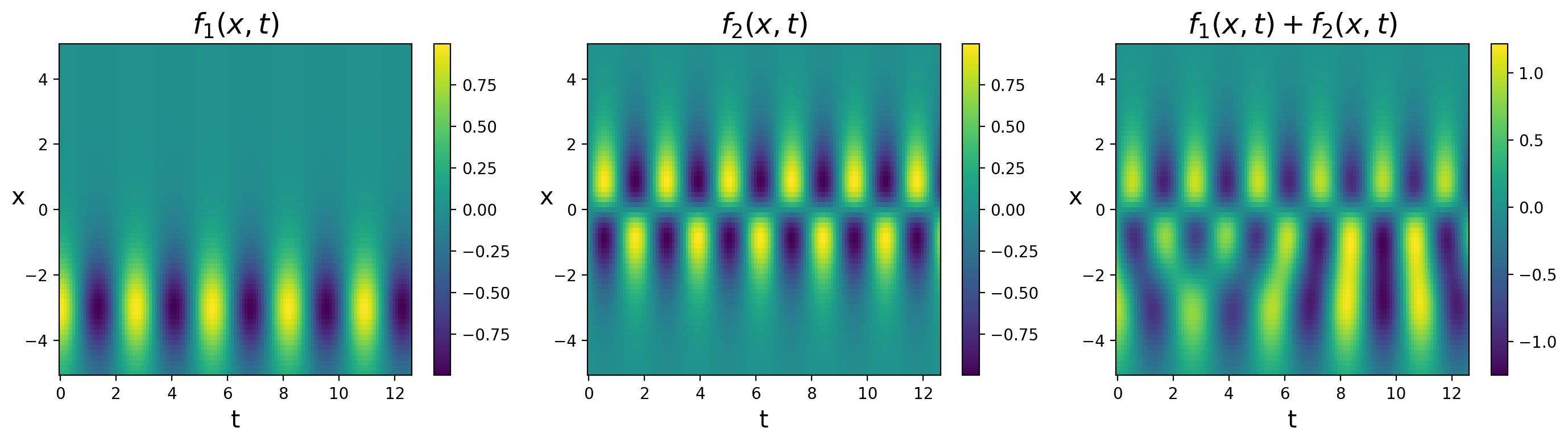}
    \caption{The synthetic data set described by Equation \eqref{eq:example}. The data consists of two spatiotemporal signals $f_1$ and $f_2$, and is collected along the spatial grid $x \in [-5,5]$ across times $t \in [0, 4\pi]$.}
    \label{fig:ex-data}
\end{figure}

Consider the following synthetic system, which consists of two distinct spatiotemporal signals $f_1$ and $f_2$. Notice that each signal possesses its own unique spatial signature and its own unique temporal frequency of oscillation, which we define to be $\omega_1 = 2.3$ and $\omega_2 = 2.8$ respectively.
\begin{align}
    f(x,t) &= f_1(x,t) + f_2(x,t) \nonumber \\
    &= \text{sech}(x+3)\cos(2.3t) + 2\text{sech}(x)\tanh(x)\sin(2.8t).
    \label{eq:example}
\end{align}
We specifically examine this system along the spatial grid $x \in [-5, 5]$ with dimension $n=65$ across $m=129$ uniformly-spaced time points collected from times $t \in [0, 4\pi]$. We may then arrange our snapshot data into the columns of the data matrix $\mathbf{X} \in \mathbb{R}^{65 \times 129}$. Provided below is code that may be used to generate this data set. See Figure \ref{fig:ex-data} for a visualization of the resulting data set.

\begin{lstlisting}[language=Python, numbers=none]
import numpy as np

def f1(x, t):
    return 1.0 / np.cosh(x + 3) * np.cos(2.3 * t)

def f2(x, t):
    return 2.0 / np.cosh(x) * np.tanh(x) * np.sin(2.8 * t)

nx = 65 # number of grid points along space dimension
nt = 129 # number of grid points along time dimension

# Define the space and time grid for data collection.
x = np.linspace(-5, 5, nx)
t = np.linspace(0, 4 * np.pi, nt)
xgrid, tgrid = np.meshgrid(x, t)

# Data consists of 2 spatiotemporal signals.
X1 = f1(xgrid, tgrid).T
X2 = f2(xgrid, tgrid).T
X = X1 + X2 # (65, 129) numpy.ndarray of data
\end{lstlisting}

Before we apply DMD to our data, we must first establish a few crucial observations. First, because our data consists of two distinct spatiotemporal signals, and because our data is completely real-valued, we will need $r=4$ DMD modes in order to model this data set with Equation \eqref{eq:dmd} since each of the two modes requires a complex conjugate pair. Second, although the spatial dimension of $\mathbf{X}$ far exceeds $r=4$, we find that time-delays are actually necessary if we hope to reveal the full intrinsic rank of $\mathbf{X}$. Note that this is because our data matrix is completely real-valued, and because the true spatial modes of our system do not shift in space. Though even without this knowledge, one can easily observe this via the rank of $\mathbf{X}$, as it increases from two to four after the use of any number of time-delays.

\begin{figure}
    \centering
    \includegraphics[scale=0.54]{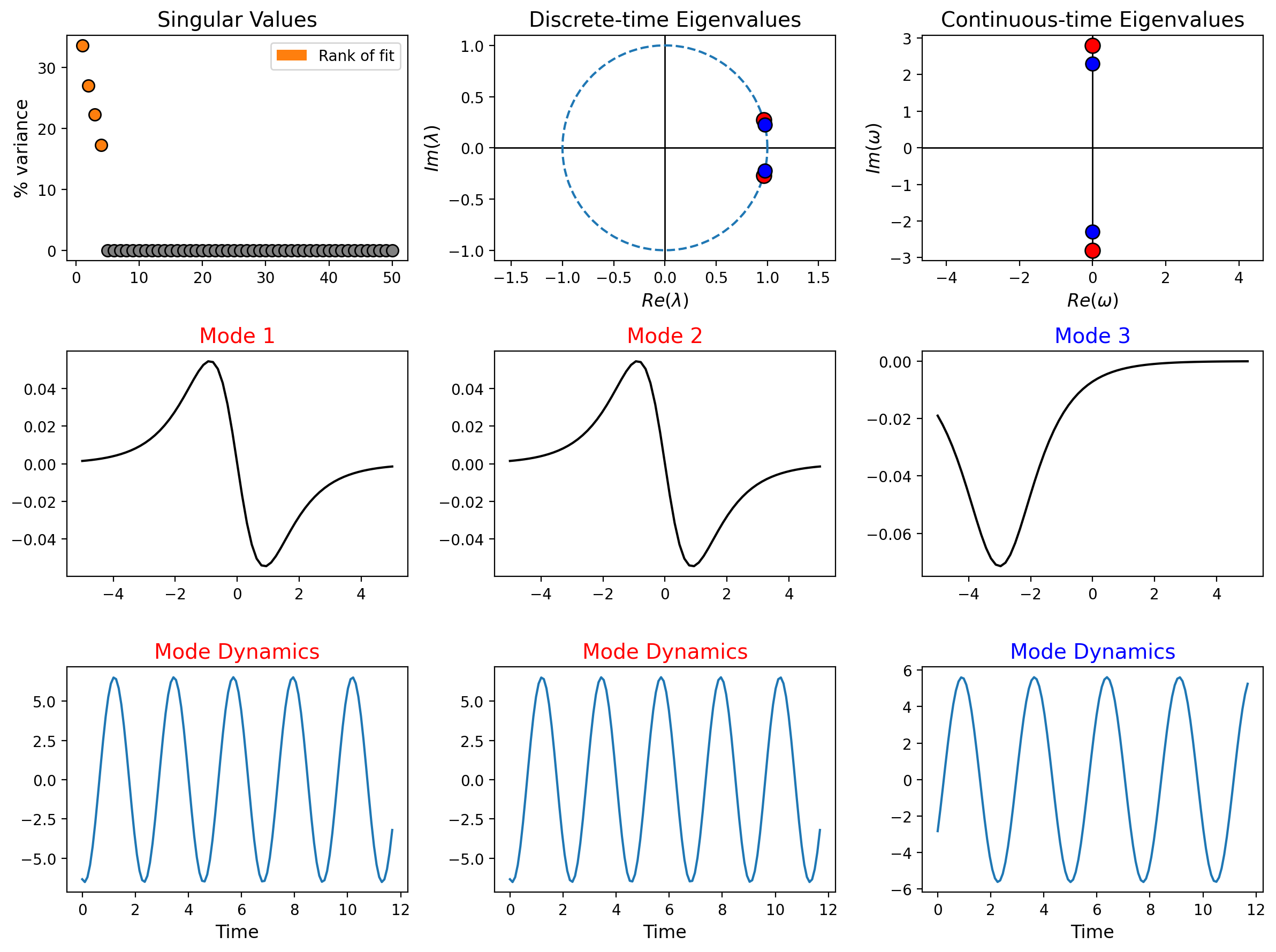}
    \caption{\texttt{plot\_summary} output using a \texttt{PyDMD} model fitted to the synthetic data in Figure \ref{fig:ex-data}. The function always produces a $3 \times 3$ grid that plots (1) the singular value spectrum of the data matrix $\mathbf{X}$, (2) the ``discrete-time eigenvalues" $\lambda_i$ of the linear operator $\mathbf{A}$, (3) the ``continuous-time eigenvalues" from $\bm{\omega}$, (4-6) three spatial modes from $\bm{\Phi}$, which default to the modes with the highest corresponding amplitudes in $\mathbf{b}$, and (7-9) the time dynamics from $\mathbf{T}(\bm{\omega})$ that are associated with the plotted modes. Associations between eigenvalues, modes, and dynamics are indicated via color coordination, and the size of an eigenvalue marker reflects the amplitude, or the general significance, of that DMD eigenvalue, mode pairing. See Section \ref{sec:background} for notation definitions.}
    \label{fig:ex-summary}
\end{figure}

Hence in order to apply DMD to our data, we first import and and initialize the \texttt{PyDMD} module that corresponds with our DMD method of choice. Since our data is evenly-spaced, noise-free, and sufficiently high-dimensional after the use of time-delays, we opt for exact DMD as per the advice of Figure \ref{fig:dmd-flow}, which is implemented by the \texttt{DMD} module. Our desired rank $r=4$ can be enforced via the \texttt{svd\_rank} parameter. Next, in order to utilize time-delays, we simply wrap our \texttt{DMD} instance in the \texttt{hankel\_preprocessing} tool, which we note is one of several data preprocessors that can now be found in the \texttt{pydmd.preprocessing} suite. Here we use $d=10$ delays for demonstration purposes, however many other $d$ values yields similar results. Finally, we invoke our \texttt{DMD} instance's \texttt{fit} method and pass in our data matrix $\mathbf{X}$ to perform the DMD pipeline.

After fitting our \texttt{PyDMD} model, we then gain access to a variety of DMD diagnostics. Although this information can often be accessed directly from our fitted model, the \texttt{PyDMD} package comes equipped with several convenient visualization tools found in \texttt{pydmd.plotter}. For standard DMD analyses, we recommend using the \texttt{plot\_summary} routine, which takes a fitted \texttt{PyDMD} module along with various plotting parameters in order to plot the most prominent results of the DMD algorithm. The entirety of this analysis is performed in the following code snippet.
\begin{lstlisting}[language=Python, numbers=none]
from pydmd import DMD
from pydmd.plotter import plot_summary
from pydmd.preprocessing import hankel_preprocessing

dmd = DMD(svd_rank=4)
delay_dmd = hankel_preprocessing(dmd, d=10)
delay_dmd.fit(X)
plot_summary(delay_dmd, x=x, t=t[1]-t[0], d=10)
\end{lstlisting}

The plot that results from this analysis is provided in Figure \ref{fig:ex-summary}. Notice that as expected, our data matrix possesses rank $r=4$ after we apply time-delays, as indicated by the singular value spectrum. Also notice how DMD successfully recovers two unique DMD modes, where each mode captures the spatial signature of either $f_1$ or $f_2$, and corresponds with a complex conjugate pair of DMD eigenvalues that captures the correct corresponding frequency of oscillation.

\subsection{Building complex PyDMD models}
\label{sec:example-2}
In the previous example, we assume access to perfectly clean, ideal simulation data that readily lends itself to exact DMD once time-delay preprocessing has been performed. However for most real-world applications of DMD, the use of some optimization or methodological extension of exact DMD is crucial or even necessary for obtaining robust models. Luckily, doing so with \texttt{PyDMD} is quite easy, as different DMD methods can be utilized simply by leveraging the appropriate \texttt{PyDMD} module.

In the code snippet below, we demonstrate how one might deploy the DMD analysis described in Section \ref{sec:example-1}, but with the BOP-DMD algorithm instead of the exact DMD algorithm. The main adjustment to our pipeline involves defining, wrapping, and fitting an instantiation of the \texttt{BOPDMD} class rather than the \texttt{DMD} class. Notice that the \texttt{BOPDMD} module possesses a variety of unique parameters that are specific to that particular algorithm, as is the case for essentially all modules of \texttt{PyDMD}. For example, in addition to the rank $r$, the number of bagging trials and the amount of data to use per trial may be defined. Users may even constrain the structure of the BOP-DMD eigenvalues for added robustness to noise, or redefine the parameters of the variable projection routine. Below, we specifically constrain the eigenvalues $\bm{\omega}$ to be purely imaginary and present with their complex conjugate. We also alter the variable projection tolerance and turn on verbosity to track variable projection progress.

\begin{lstlisting}[language=Python, numbers=none]
from pydmd import BOPDMD

bopdmd = BOPDMD(
    svd_rank=4,
    num_trials=100, trial_size=0.8,
    eig_constraints={"imag", "conjugate_pairs"},
    varpro_opts_dict={"tol":0.001, "verbose":True},
)
d = 10 # Re-apply the time-delay pipeline, again with d=10.
delay_bopdmd = hankel_preprocessing(bopdmd, d=d)
delay_t = t[:-d+1] # Time vector is truncated due to delays.
delay_bopdmd.fit(X, t=delay_t) # BOPDMD needs snapshots and times.
plot_summary(delay_bopdmd, x=x, t=t[1]-t[0], d=d)
\end{lstlisting}

The \texttt{BOPDMD} module, as well as all \texttt{PyDMD} modules in general, possess a plethora of tunable parameters not featured here, which is why we direct readers to the \href{https://pydmd.github.io/PyDMD/index.html}{{\color{blue}\texttt{PyDMD} documentation}} or simply to our \href{https://github.com/PyDMD/PyDMD}{{\color{blue}source code}} for exhaustive descriptions of all available modules. We also direct users to \href{https://github.com/PyDMD/PyDMD/blob/master/tutorials/tutorial1/tutorial-1-dmd.ipynb}{{\color{blue}Tutorial 1}} for an even greater in-depth analysis of the synthetic system given by Equation \eqref{eq:example}.

\section{Practical Tips}
In this section, we provide a recap of our most crucial tips for practical DMD usage.

\begin{itemize}
    \item \textbf{Avoid exact DMD for most applications.}

    Despite often being viewed as the standard DMD algorithm, exact DMD tends to be quite limited when it comes to real-world applications of DMD. This is because the method only permits relatively ideal data, which is often unavailable in a real-world setting. More specifically, measurements must be evenly-spaced in time, low-noise, sufficiently high-dimensional, and well-modeled with Equation \eqref{eq:linear} if exact DMD is to succeed. Some of these issues are mitigated by specific methodological extensions of DMD, in which case we recommend that they are utilized when necessary. However when it comes to standard DMD applications with real-world data, we recommend that users opt for the more robust BOP-DMD formulation~\cite{bopdmd}.

    \item \textbf{Use time-delays and inspect the rank of your data matrix.}

    Time-delay coordinates are a powerful tool that help us to unveil hidden or latent variables from data that is available to us. These extra variables then have the potential to aid our DMD modeling capabilities, as was demonstrated in Section \ref{sec:examples}. In practice, we can often never be certain that our data matrix possesses a sufficiently high rank. Perhaps unbeknownst to us, our data lacks some crucial observables, that or perhaps the underlying structure of our system requires the use of time-delays like in the case of in Section \ref{sec:examples}. For this reason, we recommend that users always inspect the rank of their data matrix and observe how time-delays impact this rank. For more information on time-delay embeddings, we refer readers to foundational works in time-delay embedding theory \cite{takens, embedology}, as well as to a few crucial works that examine the intersection of time-delays and the DMD approach \cite{havok,hankeldmd}.

    \item \textbf{Utilize your expert knowledge of the data to pick a DMD method.}

    If your data is severely polluted by noise, but you know that the underlying modes of your system should oscillate in time, use BOP-DMD with eigenvalue constraints to enforce these oscillations. If you know that your system must obey a physical principle, use physics-informed DMD to enforce it. If you know that your system is impacted by forcing, that is is multiscale, or that it is parameterized, and if you want to account for these properties in your models, use an appropriate methodological extension as opposed to standard DMD. Again, many DMD variants draw their strengths from unique algorithmic formulations that lend themselves to particular problems and data sets. We hence advise users to exploit any knowledge that they might have about their data in order to choose the method that's best for them. When in doubt, users can always refer to Figures \ref{fig:pydmd-overview} and \ref{fig:dmd-flow}.

    \item \textbf{Monitor the quality of your DMD models and explore necessary alternatives.} 

    We may not always pick the right DMD method on our first attempt, and that is okay too. Perhaps unbeknownst to us, our system of interest is highly nonlinear and it simply cannot be modeled by Equation \eqref{eq:linear}, in which case LANDO \cite{lando} might be the best algorithm choice. That or perhaps our data contains transient or multiscale features that were not properly detected and accounted for, in which case multiresolution CoSTS \cite{sliding_dmd} might the best option. In general, it is never a bad idea to start with the standard DMD approach and to apply BOP-DMD, especially if you are unsure if your data needs a special DMD variant. However with that in mind, it is crucial to monitor the accuracy of your DMD models and to fall back on alternative methods and methodological extensions when necessary.
\end{itemize}

\section{Conclusion}
The \texttt{PyDMD} package in an open-source project that enables users with diverse backgrounds in mathematics to apply DMD within a user-friendly Pythonic environment. Our latest updates featured in \texttt{PyDMD} version 1.0 additionally make it easier than ever for users to apply, and visualize results from, state-of-the-art DMD methods that are capable of extracting coherent spatiotemporal structures from real-world data sets. We hope that through this work and through future works like this, \texttt{PyDMD} can continue to serve as a practical data analysis tool and as an ever-expanding centralized code base for DMD methods, both old and new.

\section{Acknowledgements}
We wish to acknowledge the support of the National Science Foundation AI Institute in Dynamic Systems grant 2112085 (S.M.I, S.L.B. and J.N.K.). This work has been partially supported by the consortium iNEST (Interconnected North-East Innovation Ecosystem), Piano Nazionale di Ripresa e Resilienza (PNRR) – Missione 4 Componente 2, Investimento 1.5 – D.D. 1058 23/06/2022, ECS00000043, supported by the European Union's NextGenerationEU program, and by European Union Funding for Research and Innovation --- Horizon Europe Program --- in the framework of European Research Council Executive Agency: ERC POC 2022 ARGOS project 101069319 ``Advanced Reduced order modellinG: Online computational web server for complex parametric Systems'' P.I. Professor Gianluigi Rozza.

\newpage
\section*{Annotated Bibliography}
\begin{itemize}
    \item General DMD references:
    \begin{itemize}
        \item Foundational works \cite{schmid_2010, tu_2014, dmd_book, schmid2022dynamic}
        \item Connections to the Koopman operator \cite{rowley_2009, tu_2014, modern_koopman} 
        \item Studies on the effects of noise \cite{bagheri_2014, noise_1, noise_2, fbdmd, tdmd}
    \end{itemize}

    \item DMD application areas:
    \begin{itemize}
        \item Fluid Dynamics \cite{schmid_2009, schmid_2010, schmid_2008, Noack2016jfm}
        \item Epidemiology \cite{proctor_2015} 
        \item Neuroscience \cite{brunton_2016, alfatlawi_2020} 
        \item Finance \cite{mann_2016} 
        \item Plasma Physics \cite{taylor_2018, kaptanoglu_2020} 
        \item Video Processing \cite{grosek_2014} 
        \item Robotics \cite{berger_2015, abraham_2019, bruder_2019} 
        \item Power Grids \cite{sinha_2020, susuki_1, susuki_2}
    \end{itemize}

    \item \texttt{PyDMD} features prior to version 1.0:
    \begin{itemize}
        \item Exact DMD \cite{tu_2014}
        \item DMD with control \cite{dmdc} 
        \item Multiresolution DMD \cite{mrdmd} 
        \item Sparsity-promoting DMD \cite{spdmd} 
        \item Compressed DMD \cite{cdmd} 
        \item Time delay DMD \cite{hankeldmd} 
        \item Higher order DMD \cite{hodmd} 
        \item Forward-backward DMD \cite{fbdmd} 
        \item Total least-squares DMD \cite{tdmd} 
        \item Optimal closed-form DMD \cite{not_optdmd} 
        \item Subspace DMD \cite{subspacedmd}
        \item Original \texttt{PyDMD} paper \cite{pydmd}
    \end{itemize}

    \item New \texttt{PyDMD} features as of version 1.0:
    \begin{itemize}
        \item Optimized DMD \cite{optdmd} 
        \item Bagging, optimized DMD (BOP-DMD) \cite{bopdmd} 
        \item Coherent spatiotemporal scale separation (CoSTS) \cite{sliding_dmd} 
        \item Parametric DMD \cite{paradmd, paradmd_2} 
        \item Randomized DMD \cite{rdmd} 
        \item Physics-informed DMD \cite{pidmd} 
        \item Extended DMD \cite{edmd, edmd_kern, edmd_2} 
        \item Hankel alternative view of Koopman (HAVOK) \cite{havok, shavok} 
        \item Linear and nonlinear disambiguation optimization (LANDO) \cite{lando} 
        \item DMD with centering \cite{centering}
    \end{itemize}

\end{itemize}

\newpage
\bibliographystyle{unsrtsiam}
\bibliography{refs}

\end{document}